\documentclass{edp-conf}
\usepackage{graphicx}
\begin{document}

\TitreGlobal{SF2A 2005}

\title{On Artificial Transits Feasibility and SETI}
\author{Arnold, Luc}\address{OHP CNRS 04870 Saint-Michel-l'Observatoire, France (arnold at obs-hp.fr)}
\runningtitle{Artificial Transits and SETI}
\setcounter{page}{237}
\index{Arnold, Luc}

\maketitle
\begin{abstract} 
It is known that the shape of a planet (oblateness, rings, etc.) slightly modifies the shape of the transit light curve. The forthcoming space missions (Corot, Kepler), able to detect the transit of Earth-like planets, could a fortiori also detect the transit of artificial planet-size objects if their shape is significantly different from a natural (planetary) object. Multiple artificial objects would also produce transit light curves easily recognizable from natural transits.  Artificial transits, especially of multiple objects, could be used for the transmission of clear attention-getting signals, with a sky coverage (efficiency) comparable to that of the laser pulse method. Although out of reach of current human technologies, the building of an Earth-size $1 \mu m$ thick mask would require energy and bulk material amounts already managed on Earth today. The migration of the mask toward an inner orbit and its protection against asteroids or meteoroids are also briefly discussed.
\end{abstract}
%
\section{Introduction: stellar communication with artificial transits}
It is already known that the shape of a planet (its oblateness or the presence of rings, etc.) slightly modifies the shape of the transit light curve (Barnes \& Fortney 2003, 2004). In a recent paper (Arnold 2005), we have analyzed the transit light-curve signatures of artificial (i.e. non-planetary) planet-size objects, possibly built by an extraterrestrial advanced civilization. We have shown that an artificial object leaves a specific signature in the light-curve depending on its shape. The object shape can be chosen (rotating triangle, screen with holes, etc.) to produce a signature recognizable from a natural transit. 

An artificial transit, considered as an elementary bit of information, is visible (transmitted)  in a relatively large solid angle at a given time, but requires a complete orbit to be observed a second times. If we consider stellar communication with laser (Kingsley 2001), bits of information (laser pulses) can be much more numerous but only sent in a narrow solid angle. Calculation shows that the amount of bits transmitted through a given solid angle per unit of time are similar for both methods. Moreover both have roughly the same range with our assumptions.  

For a Jupiter-size object, the signature is in the $10^{-4}$ magnitude range, thus within the photometric resolution of Corot or Kepler missions. For Earth-size objects, it is not possible to detect details in the extremely shallow ($\approx 10^{-4}$) transit light-curve. To produce a distinguishable transit signal with objects as small as the Earth, a solution could be to arrange transits of multiple objects in a particular sequence such as prime numbers. 

\section{Feasibility}
Let's consider the building of a $1\mu m$ thick, $12000km$ in diameter mask made of iron. The required volume of iron represents a sphere of $632m$ in diameter.
It is amazing to see that its mass, $1.04 \times 10^{12} kg$, is almost exactly the mass of iron (steel) produced in the world in 2004, just above one billion-tonne \footnote{International Iron \& Steel Institute 2005 report at $http://www.worldsteel.org/wsif.php$}.
The energy required to heat this amount of iron from $\approx 0$ to $1808 K$  is $7.1\times  10^{14} Wh$, and its fusion at $1808 K$ needs  
$2.3\times  10^{14}$ more $Wh$. This represents 3 days of the 2002 world energy mean daily consumption\footnote{Energy Information Administration report DOE/EIA-0484(2005)  July 2005 $http://www.eia.doe.gov/oiaf/ieo/$}, obviously consistent with the annual production of steel mentioned above. These numbers suggest that the building of such a large mask, possibly from material collected on an iron-rich M-type asteroid, might be within human technology capabilities in the future.

Assuming the mask is built in the Main Asteroid Belt, it would be necessary to transfer it toward the inner solar system, in order to increase the solid angle from which
the transit is visible and thus increase the communication efficiency. This migration is possible in principle through solar sailing: the mask can be oriented
 to be decelerated by the pressure of solar radiations, inducing its controlled fall toward an orbit closer to the Sun.

To protect the mask against asteroids or meteoroids encounter in the Main Belt, it can be produced into small \textit{sails} and then migrated into a safer inner orbit where all sails are assembled. These much smaller parts might be connected together by mechanical fuses adjusted to break
above a given acceleration. A bolide colliding with a strength above the fuses threshold would take with it the impacted part from the rest of the mask, producing a small hole in the structure. This at least seems applicable for non- or slowly-rotating masks, because the acceleration at the outer edge of the mask requires stronger thus less sensitive fuses.


\end{document}